\documentclass[12pt]{article}%
\usepackage[T2A]{fontenc}
\usepackage[cp1251]{inputenc}
\usepackage{amsmath}
\usepackage{amsfonts}
\usepackage{amssymb}
\usepackage{graphicx}
\usepackage{cite}%
\begin{document}

\title{Stable components of sound fields in the ocean}
\author{A.L. Virovlyansky\\
\textit{{\small Institute of Applied Physics, Russian Academy of Science}}\\
\textit{{\small 46 Ul'yanov Street, 603950 Nizhny Novgorod,
Russia}} \\
\date{}} \maketitle

\bigskip

\newpage

\begin{abstract}
A method is proposed for finding the wave field components which
are weakly sensitive to the sound speed perturbation in the ocean
acoustic waveguides. Such a component is formed by a narrow beam
of rays whose spread in vertical direction,  up to the observation
range, remains less than the vertical scale of perturbation. These
rays pass through practically the same inhomogeneities and
therefore their phases acquire close increments. If the ray
amplitudes vary insignificantly, then (i) the stable component of
the monochromatic field in the perturbed and unperturbed waveguide
differ by only a constant phase factor, and (ii) in the case of
transient wave field the perturbation causes only an additional
time delay of the stable component as a whole. It is shown how the
stable components can be selected from the total wave field using
the field expansions in coherent states or in normal modes. The
existence of stable components is demonstrated by numerical
simulation of sound field in a deep water waveguide. It turns out,
that even though the assumptions (i) and (ii) are not met exactly,
the stable components in the perturbed and unperturbed waveguide
are quite close.

PACS number(s): 43.30.Cq, 43.30.Bp, 43.30.Re

\end{abstract}

\bigskip

\newpage

\section{Introduction}

The main factor reducing the accuracy of field simulation in the
ocean acoustic waveguide is the inevitable inaccuracy of
environmental model used in solving the wave equation. When
analyzing this issue, the main attention is usually paid to
questions on how the uncertainty in environmental parameters
manifests itself in field predictions
\cite{Dowling2008,LePage2006} and in solutions of inverse problems
\cite{Gerstoft1998,Lin2006}. This work is devoted to another
aspect of the problem of inaccurate knowledge of the environment
model. We argue that at relatively short ranges there exist such
components of the wave field which are weakly sensitive to the
variations of waveguide parameters and can be predicted with a
reasonable accuracy even under conditions of uncertain
environment.

For the analysis of field variations caused by a weak sound speed
perturbation we use the geometrical optics approximation. The
geometrical-optics description takes especially simple form at
short ranges where the deviation of ray path from its unperturbed
position caused by the perturbation is negligible
\cite{BL2003,FDMWZ79}. At these range the ray amplitude remains
almost the same and its phase acquires an increment $\phi$, which,
generally, is not small and can exceed $\pi$. The object of our
study is a component of the total field formed by a narrow beam of
rays escaping a point source at close launch angles. As long as
the spread of ray paths along the vertical coordinate remains
small compared to the vertical scale of perturbation all the rays
pass through practically the same inhomogeneities and their phase
increments vary synchronously. At ranges where the phase
increments $\phi$ of these rays are almost the same, the beam
forms a component of the total field which we call stable. Its
stability manifests itself as follows. (i) In the presence of
perturbation this component is multiplied by a phase factor
$e^{i\phi}$, where $\phi$ (it is not assumed to be small) depends
on the perturbation and range but does not depend on the depth.
(ii) If the source emits a pulse signal, then this beam of rays
forms a stable component of the transient field. In the presence
of perturbation this component acquires only an additional
depth-independent time delay.

The so-determined stable component of the total field at the range
of observation is formed by rays arriving at close grazing angles
into a small depth interval. In terms of the Hamiltonian formalism
these rays arrive at a small area of the phase plane. We say that
the stable component is associated with this area.

Our objective is to extract the stable component from the total
field. To solve this problem, two methods are proposed. One of
them is based on using the coherent state formalism borrowed from
quantum mechanics. Another method is based on the modal
representation of the wave field. The use of these instruments is
explained by the fact that the coherent state and normal mode
describe the components of the wave field formed by contributions
of rays from small cells of the phase plane. These cells have
different shapes but equal areas.

The existence of stable components is demonstrated by numerical
simulation on an example of a deep-water waveguide. We compare
sound field components formed by the same beam of rays in the
unperturbed waveguide and in the presence of sound speed
fluctuations. Their closeness is characterized by a quantitative
criterium. It turns out that even though the conditions (i) and
(ii) are not met exactly the stable components in the perturbed
and unperturbed waveguide are quite close.

The paper is organized as follows. Heuristic arguments on the
existence of stable components of the sound field formed by narrow
beams of rays are presented in Sec. \ref{sec:heuristic}. Section
\ref{sec:env} describes a model of the waveguide, which is then
used to illustrate and test the results. Procedures for
construction of stable components from coherent states and normal
modes are formulated in Secs. \ref{sec:coh} and \ref{sec:modes},
respectively. Section \ref{sec:numeric} presents the numerical
evidences confirming the existence of stable components.  In Sec.
\ref{sec:concl} the results of this work are summarized.

\section{Heuristic arguments for the existence of stable components
\label{sec:heuristic}}

Neglecting horizontal refraction, we consider a sound field at a
carrier frequency $f$ in a waveguide with the sound speed $c\left(
r,z\right)  $, where $r$ is the range and $z$ is the depth. The
refractive index is $n\left( r,z\right)  =c_{0}/c(r,z)$, where
$c_{0}$ is the reference sound speed satisfying condition
$\left\vert c\left(  r,z\right)  -c_{0}\right\vert \ll c_{0}$. In
what follows we assume that the sound field is excited by a
source set at point $\left(  0,z_{s}\right)  $.

In the geometrical optics approximation the wave field at the
observation point is formed by contributions from eigenrays
arriving at this point. The contribution from a single eigenray is
$A\exp\left(  ikS\right)  $, where $A$ and $S$ are the ray
amplitude and eikonal, respectively, $k=2\pi f/c_{0}$ is the
reference wave number. In the presence of a week sound speed
perturbation $\delta c\left(  r,z\right)  $, the deviation of a
ray path from its unperturbed position at short ranges can be
neglected. Then the influence of perturbation can be accounted for
by replacing $A\exp\left(  ikS\right)  $
with $A\exp\left(  ik\left(  S+\delta S\right)  \right)  $, where%
\begin{equation}
\delta S=-\frac{1}{c_{0}^{2}}\int_{\Gamma}ds~\delta c, \label{dS}%
\end{equation}
$\Gamma$ is the unperturbed ray path, and $ds$ is the arc length
\cite{BL2003,FDMWZ79}. This approximate formula is widely used in
underwater acoustics. In a deep ocean at frequencies of order 100
Hz it is applied at ranges up to a few hundred kilometers
\cite{FDMWZ79}.

In the scope of Hamiltonian formalism the ray path at range $r$ is
determined by its vertical coordinate $z$ and momentum $p=n\left(
r,z\right)  \sin\chi$, where $\chi$ is the ray grazing angle
\cite{Vbook2010}. The path is described
by functions $p\left(  r,p_{0},z_{0}\right)  $ and $z\left(  r,p_{0}%
,z_{0}\right)  $, where $p_{0}$ and $z_{0}$ are the starting momentum and
coordinate at $r=0$, respectively.

Consider a beam of ray paths leaving the source with starting
momenta in an interval $p_{0}\pm\delta p_{0}/2$. If $\delta p_{0}$
is so small that the spread of ray coordinates $z$ at any range
$r^{\prime}<r$ is small compared to the vertical scale of
perturbation $\delta c$, then it may be expected that at range $r$
phase increments $k\delta S$ of all these rays are almost the
same, that is, their differences are small compared to $\pi$. It
should be emphasized that $k\left\vert \delta S\right\vert $ is
not assumed to be small and it can exceed $\pi$. Below we will
call such rays the rays with close phase increments.

At the observation range $r$ the contributions of these rays form
a component of the total wave field whose complex amplitude in the
vertical section we denote $u_{s}(z)$. It is defined only at
arrival depths of beam rays. The component of the sound field
formed by the same rays in the unperturbed waveguide denote
$\tilde{u}_{s}\left(  z\right)  $. Since
the phase increments $k\delta S$ of all beam rays are close%
\begin{equation}
u_{s}\left(  z\right)  =e^{ik\delta S}\tilde{u}_{s}\left(  z\right)  .
\label{u-ut}%
\end{equation}

The closeness of functions $u_{s}\left(  z\right)  $ and
$\tilde{u}_{s}\left( z\right)  $ can be quantitatively
characterized by the similarity coefficient
\begin{equation}
B_{\text{cw}}\left(  u_{s},\tilde{u}_{s}\right)  =\frac{\left\vert \int
dz~u_{s}\left(  z\right)  \tilde{u}_{s}^{\ast}\left(  z\right)  \right\vert
}{\left(  \int dz~\left\vert u_{s}\left(  z\right)  \right\vert ^{2}\right)
^{1/2}\left(  \int dz~\left\vert \tilde{u}_{s}\left(  z\right)  \right\vert
^{2}\right)  ^{1/2}}, \label{B}%
\end{equation}
where the asterisk denotes complex conjugation (in quantum mechanics a similar
characteristic is called fidelity). If the depth dependence of phase increment
$k\delta S$ is weak, then the similarity coefficient $B$ is close to unity and
we call $u_{s}\left(  z\right)  $ a stable component of the total field.

If the source emits a pulse signal%
\begin{equation}
s\left(  t\right)  =\int df~g\left(  f\right)  e^{-2\pi ift}, \label{s-def}%
\end{equation}
where $t$ is the time, then the transient sound field at the observation range
$r$ is
\begin{equation}
v\left(  z,t\right)  =\int df~g\left(  f\right)  u\left(  z,f\right)  e^{-2\pi
ift}, \label{v-g}%
\end{equation}
where $u\left(  z,f\right)  $ is a monochromatic field excited by
the source at frequency $f$. Consider a beam of rays whose
increments at the observation range are close at all frequencies
within the signal bandwidth. At each of these frequencies the beam
forms a stable components $u_{s}\left(  z,f\right) $ of $u\left(
z,f\right)  $. We will call
\begin{equation}
v_{s}\left(  z,t\right)  =\int df~g\left(  f\right)  u_{s}\left(  z,f\right)
e^{-2\pi ift} \label{vs}%
\end{equation}
the stable component of the transient field. Since the eikonal
increment $\delta S$ does not depend on frequency $f$, the stable
components in the perturbed, $v_{s}$, and unperturbed,
$\tilde{v}_{s}$, waveguide are related by
equation%
\[
v_{s}\left(  z,t\right)  =\tilde{v}_{s}\left(  z,t+\delta t\right)  ,
\]
where $\delta t=\delta S/c_{0}$ is a time delay which does not depend on depth
$z$. By analogy with Eq. (\ref{B}), we introduce the similarity coefficient of
two transient wave fields%
\[
B_{\text{tr}}\left(  v_{s},\tilde{v}_{s}\right)  =\max_{\tau}\frac{\left\vert
\int dzdt~v_{s}\left(  z,t\right)  \tilde{v}_{s}^{\ast}\left(  z,t+\tau
\right)  \right\vert }{\left(  \int dzdt~\left\vert v(z,t\right\vert
^{2}\right)  ^{1/2}\left(  \int dzdt~\left\vert \tilde{v}(z,t\right\vert
^{2}\right)  ^{1/2}}%
\]%
\begin{equation}
=\max_{\tau}\frac{\left\vert \int dzdf~u_{s}\left(  z,f\right)  \tilde{u}%
_{s}^{\ast}\left(  z,f\right)  \left\vert g\left(  f\right)  \right\vert
^{2}e^{2\pi if\tau}\right\vert }{\left(  \int dzdf~\left\vert u(z,f\right\vert
^{2}\left\vert g\left(  f\right)  \right\vert ^{2}\right)  ^{1/2}\left(  \int
dzdf~\left\vert \tilde{u}(z,f\right\vert ^{2}\left\vert g\left(  f\right)
\right\vert ^{2}\right)  ^{1/2}}, \label{Bt}%
\end{equation}
where $\tau$ compensates for the unknown time delay $\delta t$.

The above arguments suggest that a narrow beam of rays with close phase
increments forms a stable component of both CW and transient wave fields.
Similarity coefficients (\ref{B}) and (\ref{Bt}) of stable components
associated with the same beam of rays in the perturbed and unperturbed
waveguide are close to unity. Below we consider two methods of extracting
stable components from the total wave field measured in an experiment or
computed in a full wave numerical simulation. Application of these methods
will be illustrated using the simulation of sound fields in a deep water
waveguide whose model is described in the next section.

\section{Environmental model for numerical simulation \label{sec:env}}

\begin{figure}[ptb]
\begin{center}
\includegraphics[
trim=0.000000in 0.000000in 0.007069in 0.006918in, height=7.5212cm,
width=6.9875cm ]{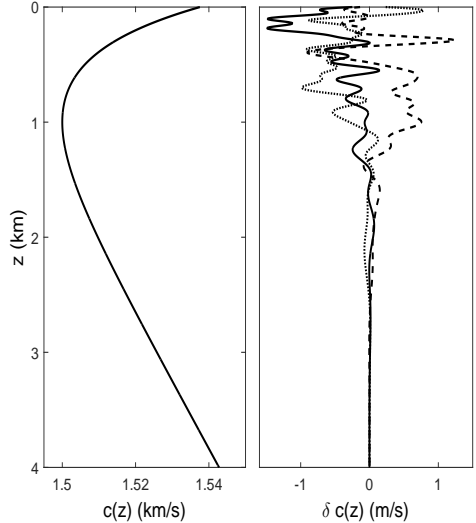}
\end{center}
\caption{Unperturbed sound speed profile (left panel) and
perturbation $\delta c$ in vertical sections of the waveguide at three
different ranges (right panel).}%
\label{fig1}%
\end{figure}

In numerical simulations presented below we use an environmental model with an
unperturbed sound speed profile representing the canonical (or Munk) profile
\cite{BL2003,FDMWZ79}%
\begin{equation}
\bar{c}(z)=c_{r}\left[  1+\varepsilon\left(  e^{\eta}-\eta-1\right)  \right]
,\;\;\eta=2(z_{a}-z)/B \label{Munk}%
\end{equation}
with parameters $c_{r}=1.5$ km/s, $\varepsilon=0.0057$, $B=1$ km,
and $z_{a}=1$ km. Profile $\bar{c}(z)$ is shown in the left panel
of Fig. \ref{fig1}. The bottom depth is 5 km.

It is assumed that the weak perturbation $\delta c(r,z)$ is caused by random
internal waves with statistics determined by the empirical Garrett-Munk
spectrum \cite{FDMWZ79}. To generate realizations of a random field $\delta
c(r,z)$ we apply a numerical technique developed by J. Colosi and M. Brown
\cite{CB98}. In their model the perturbation has the form%
\begin{equation}
\delta c(r,z)=c_{r}\frac{\mu}{g}N^{2}\zeta(r,z), \label{dc-appendix}%
\end{equation}
where $g=9.8$ m/s$^{2}$ is the gravitational acceleration, $\mu=24.5$ is a
dimensionless constant, $N(z)=N_{0}\exp(-z/L)$ is a buoyancy frequency
profile, $N_{0}=2\pi/(12$ min$)=0.0087$ 1/s is a buoyancy frequency near the
surface, $L=1$ km. The random function $\zeta(r,z)$ presents
internal-wave-induced vertical displacements of a fluid parcel. Its
realizations have been computed using Eq. (19) from Ref. \cite{CB98}. We
consider an internal wave field formed by $30$ normal modes and assume its
horizontal isotropy. Components of wave number vectors in the horizontal plane
belong to the interval from $2\pi/100$ km$^{-1}$ to $2\pi/2$ km$^{-1}$. An rms
amplitude of the perturbation scales in depth like $\exp(-3z/2L)$ and its
surface-extrapolated value in our model is about $1$ m/s. Depth dependencies
of $\delta c$ at three different ranges are shown in the right panel of Fig.
\ref{fig1}.

In what follows we consider sound fields at the observation range
$r=50$ km. It is assumed that they are excited at frequencies of
about 200 Hz by a point source set at depth $z_{s}=0.9$ km. Only
waves propagating at grazing angles $\left\vert \chi\right\vert
<12.5^{\circ}$ are taken into account. They are formed by rays
with starting momenta $\left\vert p_{0}\right\vert <0.22$.

\begin{figure}[ptb]
\begin{center}
\includegraphics[
height=2.4344in, width=3.2327in ]{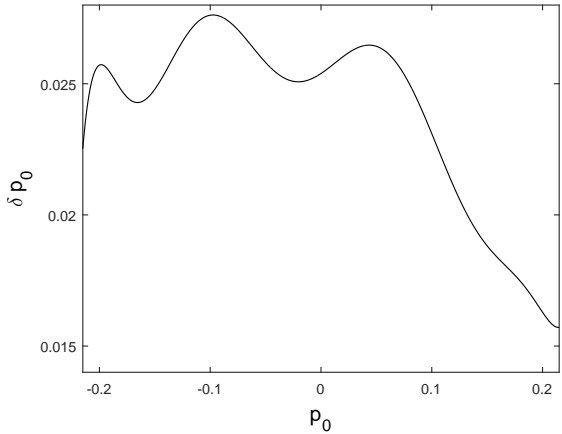}
\end{center}
\caption{Estimate of $\delta p_{0}$ defining
the interval of ray starting momenta $p_{0}\pm\delta p_{0}/2$ corresponding to
paths whose random phase increments at frequency 200 Hz remain close at 50-km range.}%
\label{fig2}%
\end{figure}

Using a standard ray tracing technique \cite{JKPS2011}, eikonal increments
$\delta S$ at $r=50$ km were computed for a dense set of $p_{0}$ in 400
realizations of random waveguide. For each $p_{0}$ an interval of momenta
$p_{0}^{\prime}$ satisfying the condition $k^{2}\left\langle \left[  \delta
S\left(  p_{0}\right)  -\delta S\left(  p_{0}^{\prime}\right)  \right]
^{2}\right\rangle <1$, where the brackets $\left\langle {}\right\rangle $
denote the averaging over realizations and $k$ is the reference wavenumber at
$f=200$ Hz, was found. The width of this interval $\delta p_{0}$ is shown in
Fig. \ref{fig2} as a function of $p_{0}$. This plot plays a crucial role in
finding stable components. For any $p_{0}$, it allows one to find an interval
$p_{0}\pm\delta p_{0}/2$ defining a beam of rays with close phase increments
at the observation range.

\section{Construction of stable components from coherent states
\label{sec:coh}}

In this section we propose a method for extracting a stable component formed
by a narrow beam of rays from the total wave field. It is based on the
formalism of coherent states widely used in quantum mechanics \cite{CM68}.

\subsection{Ray line \label{sub:ray_line}}

\begin{figure}[ptb]
\begin{center}
\includegraphics[
height=3.2673in, width=4.3526in ]{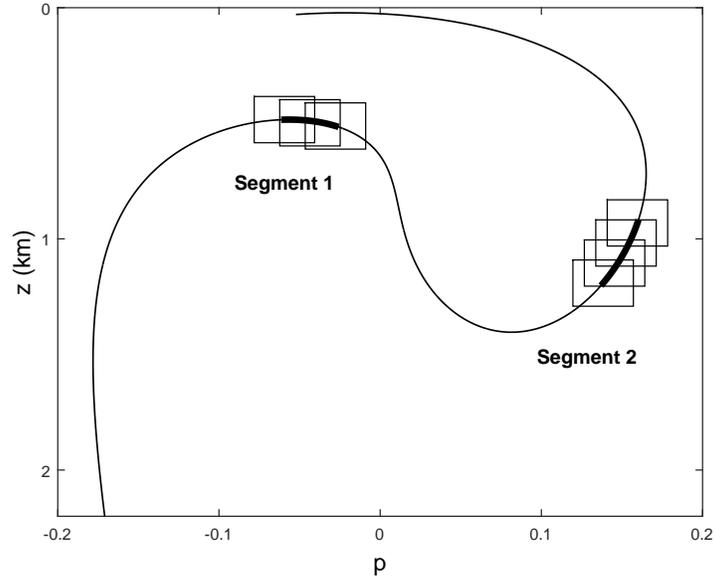}
\end{center}
\caption{Ray line at 50-km range (thin solid
line). Segments 1 and 2 are shown by thick solid lines. Areas covered by
rectangles strung on the ray line represent fuzzy versions of Segment 1 and 2.}%
\label{fig3}%
\end{figure}

Introduce the phase plane $P-Z$, where $P$ is the momentum and $Z$
is the depth. The arrival of a ray with starting momentum $p_{0}$
at the observation range $r$ is depicted in this plane by a point
with coordinate $P=p\left( r,p_{0},z_{s}\right)  $ and $Z=z\left(
r,p_{0},z_{s}\right)  $. Arrivals of rays with different $p_{0}$
form a curve which we shall call the ray line. Figure \ref{fig3}
shows the ray line at 50-km range computed for the unperturbed
waveguide described in Sec. \ref{sec:env}.

\begin{figure}[ptb]
\begin{center}
\includegraphics[
height=2.4984in, width=4.5351in ]{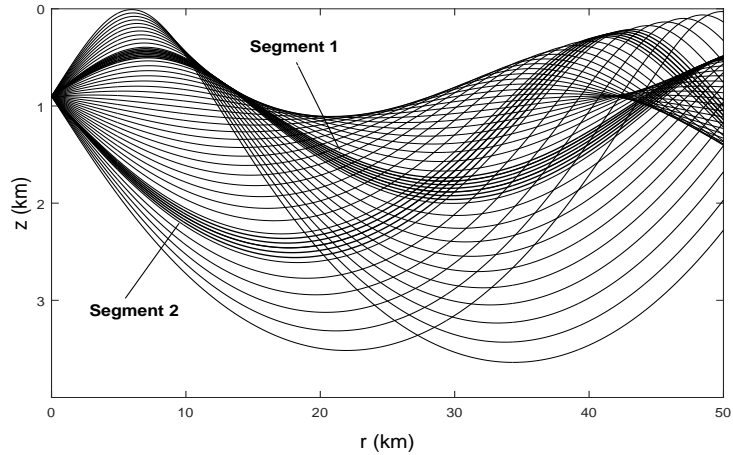}
\end{center}
\caption{Fan of rays escaping a point source set at 0.9 km. Ray
paths forming Segments 1 and 2 in Fig.
3 are plotted denser and shown by thicker solid lines, than other paths.}%
\label{fig4}%
\end{figure}

A beam of rays with starting momenta from interval $p_{0}\pm\delta
p_{0}/2$ forms a segment of the ray line. Two examples of such
segments are shown in Fig. \ref{fig3}\ by thick lines. Segments 1
and 2 are formed by rays with starting momenta close to
$p_{0}=-0.0975$ and $p_{0}=0.149$, respectively. Figure \ref{fig2}
allows one for a given $p_{0}$ to find a maximum value of $\delta
p_{0}$ for which rays forming the segment still have close phase
increments at the observation range and, hence, they still form a
stable component of the total field. For Segments 1 and 2 these
values of $\delta p_{0}$ are 0.025 and 0.018, respectively. Beams
of rays forming Segments 1 and 2 with these $\delta p_{0}$'s are
shown in Fig. \ref{fig4}.

\subsection{Coherent states \label{sub:coh}}

A coherent state associated with a point $\mu$ of the phase plane with
coordinates $\left(  P,Z\right)  $ is described by function%
\[
Y_{\mu}\left(  z\right)  =\frac{1}{\Delta_{z}}\exp\left[  ikP\left(
z-Z\right)  -\frac{\pi\left(  z-Z\right)  ^{2}}{\Delta_{z}^{2}}\right]
\]%
\begin{equation}
=\frac{1}{2\pi k}\int dp~\exp\left[  -\frac{\pi\left(  p-P\right)  ^{2}%
}{\Delta_{p}^{2}}+ikp\left(  z-Z\right)  \right]  , \label{Y-z}%
\end{equation}
where $\Delta_{z}$ and $\Delta_{p}=\lambda/\Delta_{z}$ are the scales of this
state along the axes $Z$ and $P$, respectively, $\lambda=2\pi/k$ is the
reference wavelength. Function $Y_{\mu}\left(  z\right)  $ can be interpreted
as a vertical section of a Gaussian wave beam of width $\Delta_{z}$ arriving
at the observation range at grazing angle $\chi=\arcsin P$.

In quantum mechanics wave functions $Y_{\mu}\left(  z\right)  $
represent quantum states with minimum uncertainty \cite{CM68}. In
the analogy with quantum mechanics $k^{-1}$ plays the role of
Planck constant.

Functions $Y_{\mu}\left(  z\right)  $ satisfy the condition%
\begin{equation}
C\int d\mu~Y_{\mu}\left(  z\right)  Y_{\mu}^{\ast}\left(  z^{\prime}\right)
=\delta\left(  z-z^{\prime}\right)  , \label{1-exp}%
\end{equation}
where $d\mu=dPdZ$ and $C =\sqrt{2}\Delta_{z}/\lambda$. They form a
complete system of functions and an arbitrary function $u(z)$ can
be represented in the
form of expansion%
\begin{equation}
u\left(  z\right)  =C\int d\mu~a_{\mu}Y_{\mu}\left(  z\right)  , \label{u-Y}%
\end{equation}
where%
\begin{equation}
a_{\mu}=\int dz~u\left(  z\right)  Y_{\mu}^{\ast}\left(  z\right)  .
\label{a-Y}%
\end{equation}

In should be emphasized that coherent states $Y_{\mu}$ are not orthogonal.
Their scalar product is%
\[
\int dz~Y_{\mu_{1}}\left(  z\right)  Y_{\mu_{2}}^{\ast}\left(  z\right)
=\frac{1}{\sqrt{2}\Delta_{z}}%
\]%
\begin{equation}
\times\exp\left[  -\pi\left(  \frac{\left(  P-P_{1}\right)  ^{2}}{2\Delta
_{p}^{2}}+\frac{\left(  Z-Z_{1}\right)  ^{2}}{2\Delta_{z}^{2}}\right)
+\frac{ik}{2}(P_{1}+P)\left(  Z_{1}-Z\right)  \right]  , \label{Y1Y2}%
\end{equation}
where $\mu_{1}=\left(  P_{1},Z_{1}\right)  $, $\mu_{2}=\left(  P_{2}%
,Z_{2}\right)  $. Due to the non-orthogonality of coherent states the choice
of coefficients $a_{\mu}$ in expansion (\ref{u-Y}) is ambiguous and formula
(\ref{a-Y}) used throughout the present paper gives only one of the possible options.

\subsection{Stable component associated with a segment of the ray line
\label{sub:CS}}

Function $Y_{\mu}\left(  z\right)  $ describes a component of the
total wave field associated with point $\mu$ of the phase plane.
According to Eqs. (\ref{Y-z}) and (\ref{a-Y}), it is formed mainly
by contributions from rays whose arrivals are depicted by points
of the plane $P-Z$ located in a rectangle with sizes
$\Delta_{p}\times\Delta_{z}$ centered at point $\mu$.

A component of the total field associated with a segment, we will determine by
the expression%
\begin{equation}
u_{a}\left(  z\right)  =C\int d\mu~W\left(  \mu\right)  a_{\mu}Y_{\mu}\left(
z\right)  , \label{ua-def}%
\end{equation}
where $W\left(  \mu\right)  $ is a weight function equal to unity
on the segment and rapidly decays outside it. Equation
(\ref{ua-def}) can be
rewritten in the form%
\begin{equation}
u_{a}\left(  z\right)  =\int dz^{\prime}~\Xi_{a}\left(  z,z^{\prime}\right)
u\left(  z^{\prime}\right)  , \label{us-coh}%
\end{equation}
where%
\begin{equation}
\Xi_{a}\left(  z,z^{\prime}\right)  =C\int d\mu~W\left(  \mu\right)  Y_{\mu
}\left(  z\right)  Y_{\mu}^{\ast}\left(  z^{\prime}\right)  \label{Xi-coh}%
\end{equation}
is a function which does not depend on a particular realization of $u\left(
z\right)  $ and defines a projection onto a stable component.

So defined, the stable component $u_{a}\left(  z\right)  $ is
formed mainly by contributions of rays from an area of the phase
plane covered by $\Delta _{p}\times\Delta_{z}$ rectangles strung
on the segment. We will call this area the fuzzy segment.

Component $u_{a}\left(  z\right)  $ can be evaluated for any
segment of the ray line. But according to arguments presented in
Sec. \ref{sec:heuristic}, it is stable only if the segment
includes rays with close phase increments. To satisfy this
condition, besides the smallness of interval $\delta p_{0}$ it is
necessary to select such scales $\Delta_{p}$ and $\Delta_{z}$ that
the fuzzy segment is not intersected by other parts of the ray
line. Since the scales $\Delta_{p}$ and $\Delta_{z}$ are connected
by the uncertainty relation $\Delta_{p}\Delta_{z}=\lambda$ the
latter requirement can be fulfilled only at short ranges where the
adjacent portions of the ray line are spaced far enough apart
or/and at a short enough wavelength $\lambda$.

For Segments 1 and 2 shown in Fig. \ref{fig3} the appropriate scales
at a frequency of 200 Hz can be easily selected. Figure \ref{fig3}
shows the fuzzy segments constructed for $\Delta_{z}=0.2$ km and $\Delta
_{p}=0.038$. These scales will be used in our simulation for the evaluation of
stable components.

The component of the transient wave field associated with a given segment,
$v_{a}\left(  z,t\right)  $, is synthesized from components of monochromatic
fields associated with the same segment at different frequencies $f$,
$u_{a}\left(  z,f\right)  $ (cf. Eq. (\ref{v-g})):%
\begin{equation}
v_{a}\left(  z,t\right)  =\int df~g\left(  f\right)  u_{a}\left(  z,f\right)
e^{-2\pi ift}. \label{va-ua}%
\end{equation}

\section{Construction of stable components from local eigenfunctions of the
waveguide \label{sec:modes}}

Consider an alternative method for forming the component of wave field
associated with a given segment of the ray line. It is based on the use of
mode representation of the wave field, that is, on the expansion of the wave
field in local eigenfunctions of the Sturm-Liouville problem
\cite{BL2003,JKPS2011}. Since, as in the preceding section we consider the
wave field at a fixed observation range, the argument $r$ in all functions
describing the wave field and the medium will be omitted for short.

\begin{figure}[ptb]
\begin{center}
\includegraphics[
height=2.9819in, width=3.9712in ]{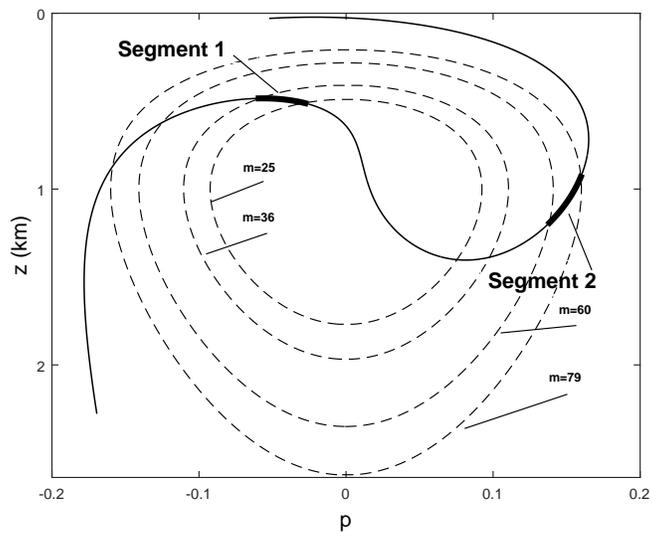}
\end{center}
\caption{Thin solid line and segments of thick solid lines show
the same ray line and Segments 1 and 2 as in Fig. 3. Dashed lines
depict the ovals corresponding to some normal modes whose numbers
are indicated next to the lines.}%
\label{fig5}%
\end{figure}

Let us exploit the expansion of a monochromatic wave field
$u\left(  z\right) $~in orthonormal local eigenfunctions of the
unperturbed waveguide $\varphi
_{m}\left(  z\right)  $:%
\begin{equation}
u\left(  z\right)  =\sum_{m=1}^{M}a_{m}\varphi_{m}\left(  z\right)
,~a_{m}=\int dz~u\left(  z\right)  \varphi_{m}\left(  z\right)  ,
\label{u-mode}%
\end{equation}
where $M$ is the number of propagating modes
\cite{BL2003,JKPS2011,BG99}. In the WKB approximation the $m$-th
eigenfunction is determined by parameters of the reference ray
whose turning depths $z_{\min}$ и $z_{\max}$ coincides with the
turning depths of the $m$-th mode. For simplicity we restrict our
attention to modes with both turning depths within the water bulk.
Momentum of the reference ray corresponding to the $m$-th mode at
depth $z~$ is
\begin{equation}
p_{m}\left(  z\right)  =\sqrt{n^{2}\left(  z\right)  -k_{m}^{2}/k^{2}},
\label{pm-def}%
\end{equation}
where $k_{m}$ is the eigenvalue of the Sturm-Liouville problem (or
the horizontal propagation constant) determined by the
quantization rule
\cite{BL2003,JKPS2011}%
\begin{equation}
k\int_{z_{\min}}^{z_{\max}}dz~p_{m}\left(  z\right)  =\pi\left(  m-1/2\right)
.\; \label{quant}%
\end{equation}
In the phase plane $P-Z$, the trajectory of reference ray represents an oval
formed by lines $P=\pm p_{m}\left(  Z\right)  $. Examples of such ovals are
shown in Fig. \ref{fig5}

The $m$-th eigenfunctions can be presented as
\cite{BL2003,Vbook2010}
\begin{equation}
\varphi_{m}(z)=\varphi_{m}^{+}(z)+\varphi_{m}^{-}(z), \label{phi-sum}%
\end{equation}
where%
\begin{equation}
\varphi_{m}^{\pm}(z)=Q_{m}\left(  z\right)  \exp\left[  \pm ik\int_{z_{\min}%
}^{z}dz^{\prime}~p_{m}\left(  z^{\prime}\right)  \mp i\pi/4\right]  ,\;
\label{phi-pm}%
\end{equation}
$Q_{m}\left(  z\right)  =\left[  D_{m}\tan\chi_{m}\left(  z\right)  \right]
^{-1/2}$, $D_{m}$ is the cycle length of the reference ray аnd $\chi
_{m}\left(  z\right)  =\arcsin(p_{m}\left(  z\right)  /n\left(  z\right)  )$
is its grazing angle at depth $z$. According to Eqs. (\ref{phi-sum}) and
(\ref{phi-pm}) the $m$-th mode represents the superposition of two Brillouin
waves whose grazing angles at horizon $z$ are $\pm\chi_{m}\left(  z\right)  $.

Take a segment of the ray line. If the oval corresponding to the
$m$-th mode intersects it at point $\left(  P,Z\right) $, then
$\left\vert P\right\vert =p_{m}\left(  Z\right)  $. It means that
a component of the total wave field formed by rays with momenta
close to $P$ at depths close to $Z$ in term of modal
representation is formed by Brillouin waves of modes with numbers
close to $m$. This fact suggests an alternative representation of
the field component associated with a given segment. We define
this component as a superposition of Brillouin waves corresponding
to a group of modes whose ovals intersects the segment. The number
of modes in the group and the number of central mode denote
$\Delta_{g}$ and $m_{c}$, respectively. Notice that the parts of
the phase plane associated with a single mode and with a single
coherent state have different shapes but the same area equal to
$\lambda$ \cite{LLquant}.

Figure \ref{fig5} shows the same ray line as in Fig. \ref{fig3}
with the same Segments 1 and 2 marked by thick lines. Dashed
curves present the ovals which pass near the endpoints of the
segments. It is seen that for Segment 1 $\Delta_{g}=12$,
$m_{c}=31$, and for Segment 2 $\Delta_{g}=20$, $m_{c}=70$.

The applicability of Eq. (\ref{phi-pm}) for the Brillouin waves is
restricted by the fact that it fails in the vicinity of mode
turning depths. But we can easily avoid the use of this analytical
expression. Our task is to sum up the components of normal modes
propagating at grazing angles corresponding to momenta $P$ of
points belonging to the segment. Such component of the $m$-th mode
denote $\Phi _{m}\left(  z\right)  $. Points of the segment cover
an interval of momenta whose width denote by $\Delta P$. The
central moment of this interval denote $P_{c}$. Function
$\Phi_{m}\left(  z\right)  $ can be found by formula
\begin{equation}
\Phi_{m}\left(  z\right)  =\frac{\lambda}{\Delta P} \int
dz^{\prime}~\varphi_{m}\left( z^{\prime }\right)  \exp\left[
ikP_{c}\left(  z-z^{\prime}\right) -\frac{\pi\left( \Delta
P\right)  ^{2}\left(  z-z^{\prime}\right)
^{2}}{\lambda^{2}}\right]
, \label{phi-mt}%
\end{equation}
which is convenient for numerical calculations. For points $z$
located far from the mode turning depths $\Phi_{m}\left(  z\right)
$ is close to $\varphi_{m}^{+}\left(  z\right)  $ or
$\varphi_{m}^{-}\left(  z\right)  $.

Based on the foregoing, we propose an alternative to Eq.
(\ref{ua-def}) expression for a field component associated with a
given segment
\begin{equation}
u_{b}\left(  z\right)  =\sum_{m}W_{m}a_{m}\Phi_{m}\left(  z\right)  ,
\label{ub-def}%
\end{equation}
where
\begin{equation}
W_{m}=\exp\left(  -\pi\frac{(m-m_{c})^{2}}{\Delta_{g}^{2}}\right),  \label{Wm}%
\end{equation}
which will be tested in numerical simulation. Although function
$u_{b}\left( z\right)  $ is formally determined at any depth $z$,
the field component is determined only within a depth interval
slightly exceeding the interval covered by the segment.

Using Eq. (\ref{u-mode}), we rewrite Eq. (\ref{ub-def}) in a form similar to
Eq. (\ref{us-coh})%
\begin{equation}
u_{b}\left(  z\right)  =\int dz^{\prime}~\Xi_{b}\left(  z,z^{\prime}\right)
u\left(  z^{\prime}\right)  , \label{us-mode}%
\end{equation}
where%
\begin{equation}
\Xi_{b}\left(  z,z^{\prime}\right)
=\sum_{m}W_{m}\Phi_{m}\left(  z\right)\varphi_{m}\left(z'\right). \label{Xi-mode}%
\end{equation}

Functions $u_{a}\left(  z\right)  $ and $u_{b}\left(  z\right)  $
describe field components formed by rays from approximately the
same area of the phase plane. Therefore it is natural to expect
that these functions should be close which means the closeness of
functions $\Xi_{a}\left(  z,z^{\prime}\right)  $ and
$\Xi_{b}\left(  z,z^{\prime}\right)  $, as well. In the next
section it will be shown in numerical simulation that this is
true.

A component of the transient wave field associated with a given
segment, $v_{b}\left(  z,t\right)  $, can be synthesized from
components of monochromatic fields at different frequencies $f$,
$u_{b}\left(  z,f\right) $, in the same manner as it is described
in Sec. \ref{sub:CS} for a component constructed from coherent
states.

\section{Numerical evaluation of stable components \label{sec:numeric}}

In this section the existence of stable components is demonstrated
in numerical simulation. We evaluate and study stable components
associated with the two segments of ray line shown in Figs.
\ref{fig3} and \ref{fig5}. The component which is constructed from
coherent states will be called the CS component, while the other
one constructed from the Brillouin waves will be called the modal
component.

\subsection{Monochromatic sound field}

Monochromatic sound field at a frequency of 200 Hz excited by a
point source set at a depth of 0.9 km was computed in 100
realizations of random waveguide describe in Sec. \ref{sec:env}.
Stable components of the total field associated with Segments 1
and 2 were evaluated in each realization. CS component, that is,
function $u_{a}\left( z\right) $, was evaluated with scales
$\Delta_{z}$ and $\Delta_{p}$ indicated in Sec. \ref{sub:CS}.
Modal component, that is, function $u_{b}\left( z\right)  $, was
evaluated with parameters $m_{c}$ and $\Delta_{g}$ indicated in
Sec. \ref{sec:modes}. Both components for each segment were found
in a depth interval from $z_{1}-\Delta_{z}/2$ to
$z_{2}+\Delta_{z}/2$, where $z_{1}$ and $z_{2}$ are minimal and
maximal depths of segment points. For Segment 1 this interval
spans from 0.38 to 0.62 km, and for Segment 2 from 0.86 to 1.3 km.

CS component is computed by formulas (\ref{us-coh}) and (\ref{Xi-coh}) with
the weight function%

\begin{equation}
W\left(  P,Z\right)  =\max_{p_{0}^{\prime}\in p_{0}\pm\delta p_{0}/2}%
\exp\left[  -4\pi\left(  \frac{\left(  P-p\left(  p_{0}^{\prime}%
,z_{s},r\right)  \right)  ^{2}}{\Delta_{p}^{2}}+\frac{\left(  Z-z\left(
p_{0}^{\prime},z_{s},r\right)  \right)  ^{2}}{\Delta_{z}^{2}}\right)  \right]
. \label{W-def}%
\end{equation}
The integration in Eq. (\ref{Xi-coh}) goes over a narrow
neighborhood of the ray line representing a part of the fuzzy
segment.

\begin{figure}[ptb]
\begin{center}
\includegraphics[
height=3.1341in, width=3.5328in ]{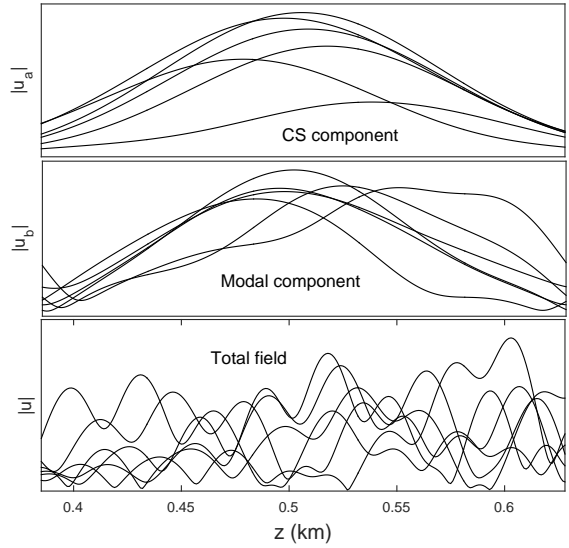}
\end{center}
\caption{Amplitudes of CS components (upper panel), modal
components (middle panel), and total fields (lower panel) for six
realizations of perturbation $\delta c$. CS and modal components
are associated with Segment 1 of the ray line.}%
\label{fig6}%
\end{figure}

Sound field excited by a point source at a frequency of 200 Hz was
computed by the method of wide angle parabolic equation
\cite{JKPS2011} for 100 realizations of perturbation $\delta c$.
Depth dependencies of the sound fields for different $\delta c$ at
the observation range 50 km are presented by functions
$u_{n}\left(  z\right)  $, $n=1,\ldots,100$. Function $u_{1}\left(
z\right)  $ describe the sound field in the unperturbed waveguide
($\delta c=0$). CS and modal components presented by functions
$u_{a,n}\left( z\right)  $ and $u_{b,n}\left(  z\right)  $,
respectively, are computed for each $u_{n}\left(  z\right)  $.
Figure \ref{fig6} presents the amplitudes of stable components
$\left\vert u_{a,n}\left(  z\right)  \right\vert $ (upper panel)
and $\left\vert u_{b,n}\left(  z\right)  \right\vert $ (middle
panel) associated with Segment 1 for six realizations of random
waveguide and the amplitude of the total field within the same
depth interval for the same realizations of the waveguide (lower
panel). The depth dependencies of amplitudes $\left\vert
u_{a,n}\left(  z\right)  \right\vert $, $\left\vert u_{b,n}\left(
z\right)  \right\vert $ and $\left\vert u_{n}\left(  z\right)
\right\vert $ for the Segment 2 look similarly (not shown).

\begin{figure}[ptb]
\begin{center}
\includegraphics[
height=2.706in, width=3.6512in ]{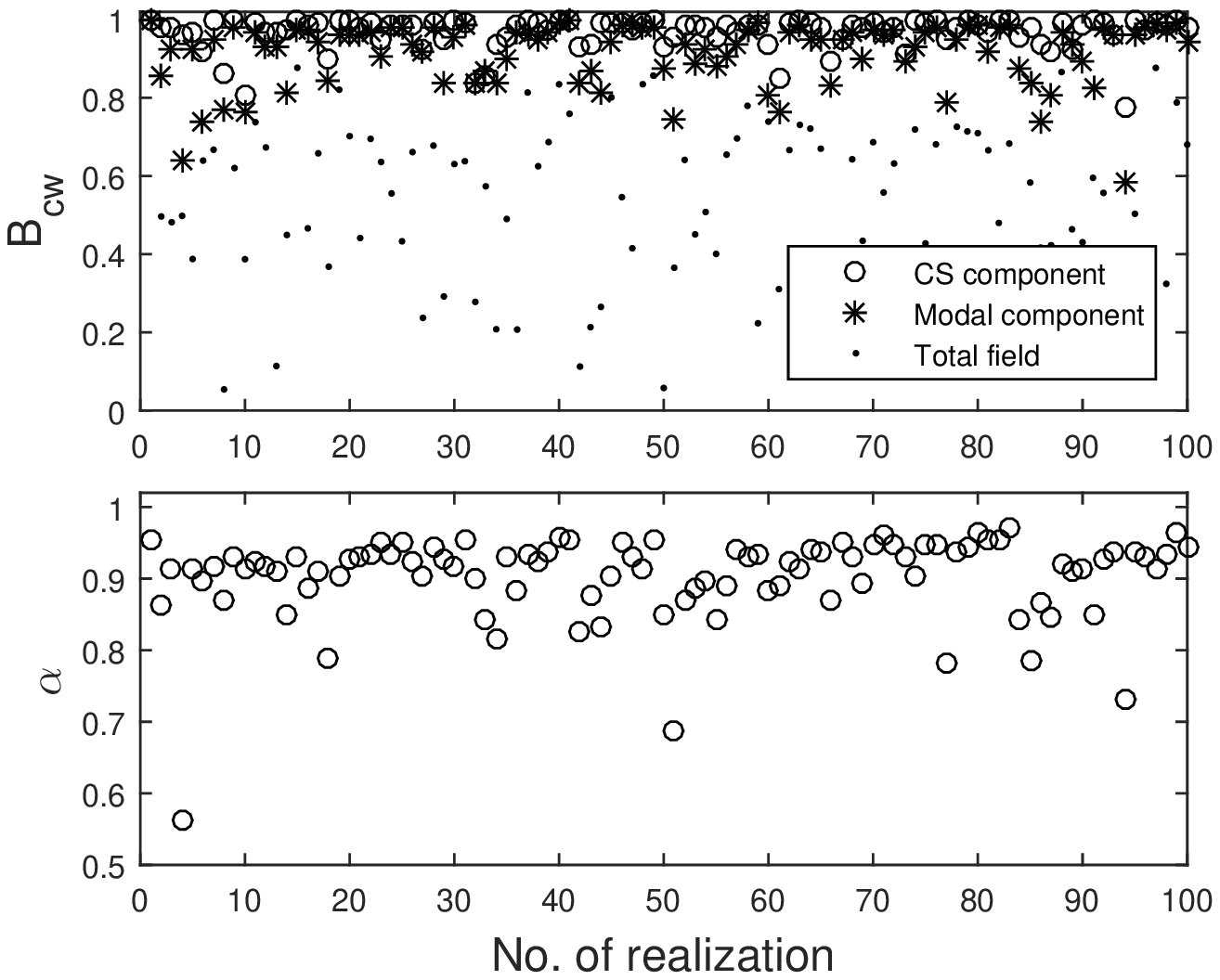}
\end{center}
\caption{Stability of field components associated with Segment 1.
Upper panel: similarity coefficient of CS components (circles),
modal component (asterisks), and total field (points) in the
perturbed and unperturbed waveguide.  Lower panel: similarity
coefficient of CS and modal
components in the same realization of random waveguide. }%
\label{fig7}%
\end{figure}

In order to quantify the stability of CS and modal components we
evaluated similarity coefficients (\ref{B}) for realizations of
these components in the perturbed and unperturbed waveguide. Upper
panel of Fig. \ref{fig7} shows similarity coefficients
$B_{cw}\left( u_{a,n},u_{a,1}\right)  $ (circles) and
$B_{cw}\left( u_{b,n},u_{b,1}\right)  $ (asterisks) for Segment 1.
Points depict similarity coefficients $B_{cw}\left(
u_{n},u_{1}\right)  $ for the total filed within the depth
interval where the stable components are defined. It is seen that
both CS and modal components are indeed more stable than the total
field.

\begin{figure}[ptb]
\begin{center}
\includegraphics[
height=2.7717in, width=3.7412in ]{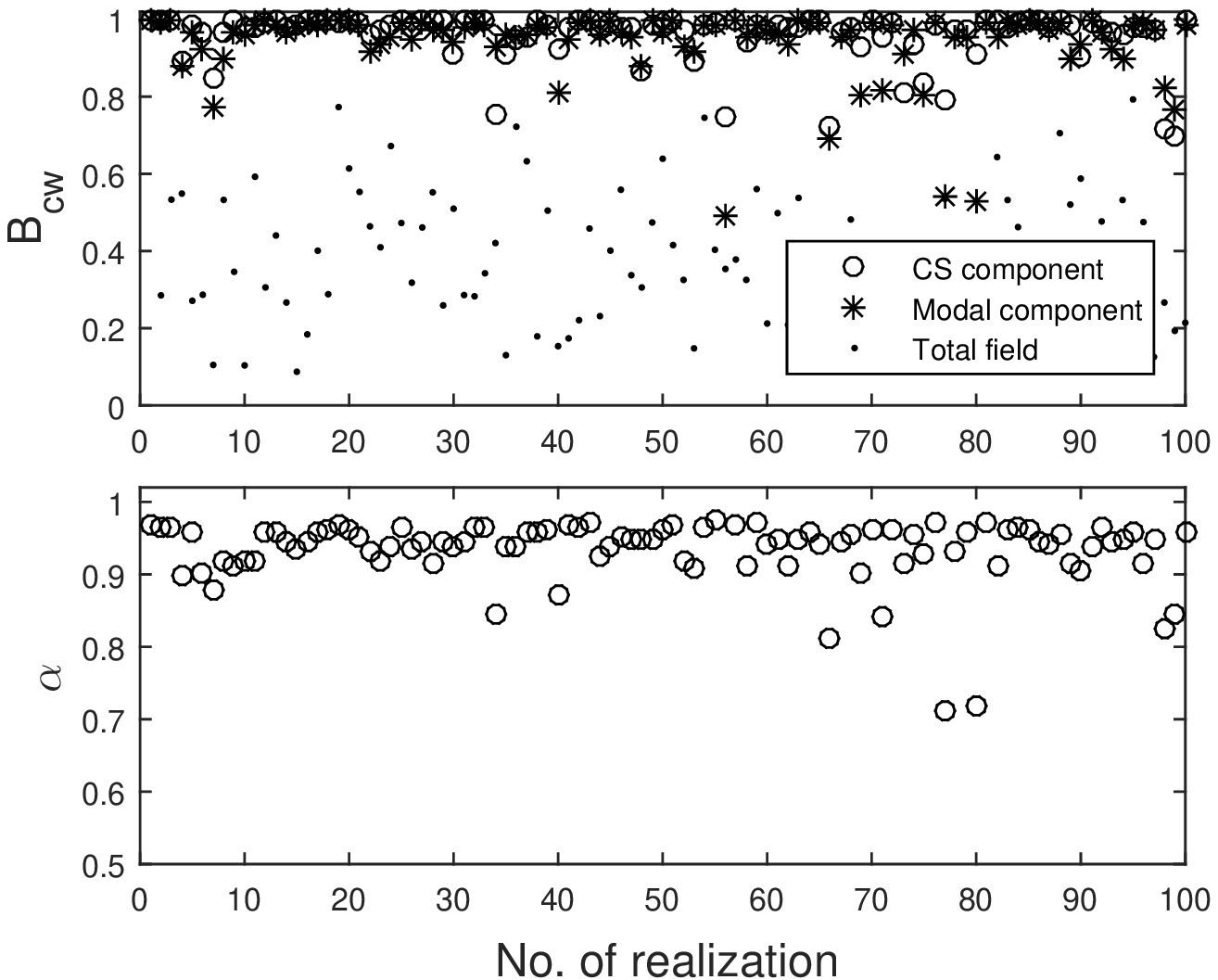}
\end{center}
\caption{The same as in Fig. 7, but for field components
associated with Segment 2.}%
\label{fig8}%
\end{figure}

To check the assumption on the closeness of CS and modal component
made in Sec. \ref{sec:modes}, the similarity coefficient
$\alpha=B\left( u_{a,n},u_{b,n}\right)  $ was computed for each
realization of perturbation. Consistent with our expectation, the
values of all these coefficients shown in the lower panel of Fig.
\ref{fig7} are rather large. Similar results are obtained for
Segment 2 (see Fig. \ref{fig8}).

\begin{figure}[ptb]
\begin{center}
\includegraphics[
height=3.0831in, width=4.0957in ]{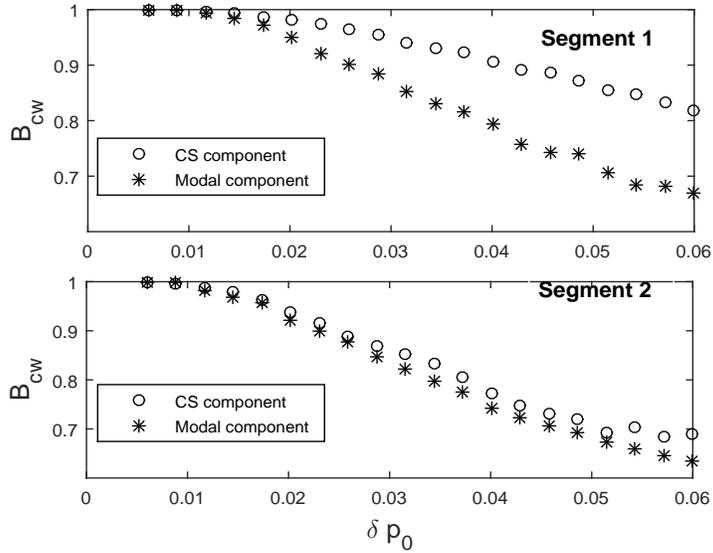}
\end{center}
\caption{Upper panel: similarity coefficients for CS (circles) and
modal (asterisks) components associated with Segment 1 in the
perturbed and unperturbed waveguide as functions of interval
$\delta p_{0}$. Lower panel: the same for components associated
with Segment 2. The similarity
coefficients are averaged over 100 realization of random waveguide.}%
\label{fig9}%
\end{figure}

As it has been indicated in Sec. \ref{sub:CS}, although CS and
modal components are defined for any parameter $\delta p_{0}$
defining the segment length, they are stable only for small enough
$\delta p_{0}$. Upper panel of Fig. \ref{fig9} presents results
illustrating the loss of stability of both CS and modal components
associated with Segment 1 at change of $\delta p_{0}$ from 0.006
to 0.06. Circles and asterisks show the similarity coefficients
$B_{cw}\left(  u_{a,n},u_{a,1}\right)  $ and $B_{cw}\left(  u_{b,n}%
,u_{b,1}\right)  $, respectively, averaged over 100 realizations
of perturbation as functions of $\delta p_{0}$. Similar results
for the CS and modal components associated with Segment 2 are
presented in the lower panel. Let us recall that according to Fig.
\ref{fig2} the maximum values of $\delta p_{0}$ for which the
field components associated with Segments 1 and 2 are expected to
be stable are about 0.025 and 0.018. This agrees with what we see
in Fig. \ref{fig9}: the decrease of similarity coefficient in the
upper panel begins at $\delta p_{0}\simeq0.02$ and in the lower
panel at $\delta p_{0}\simeq0.015$.

\subsection{Transient sound field}

Consider a point source emitting sound pulse $s\left(  t\right)
=\exp\left( -\pi t^{2}/\tau_{0}^{2}-2\pi if_{0}t\right)  $ with
$f_{0}=200$ Hz and $\tau_{0}=0.03$ s. Complex amplitude of the
sound field $v\left(  z,t\right) $ at 50-km range calculated by
formula (\ref{v-g}) with $g\left( f\right) =\exp\left(
-\pi\tau_{0}^{2}\left(  f-f_{0}\right)  ^{2}\right) $. To this
end, monochromatic fields $u_{n}\left(  z,f\right)  $,
$n=1,\ldots,22$ at 81 discrete frequencies $f$ uniformly
distributed in the interval from 160 to 240 Hz were calculated by
the method of wide angle parabolic equation for 22 realizations of
perturbation $\delta c$. These data were used to synthesize
functions $v_{n}\left(  z,t\right)  $ describing transient sound
fields in the vertical section of the waveguide at $r=50$ km.
Index $n=1$ indicates results obtained for the unperturbed
waveguide ($\delta c=0$). Then field components $v_{a,n}\left(
z,t\right)  $ and $v_{b,n}\left(  z,t\right)  $ associated with
Segments 1 and 2 were found from $v_{n}\left(  z,t\right)  $.

\begin{figure}[ptb]
\begin{center}
\includegraphics[
height=4.5in, width=4.6in ]{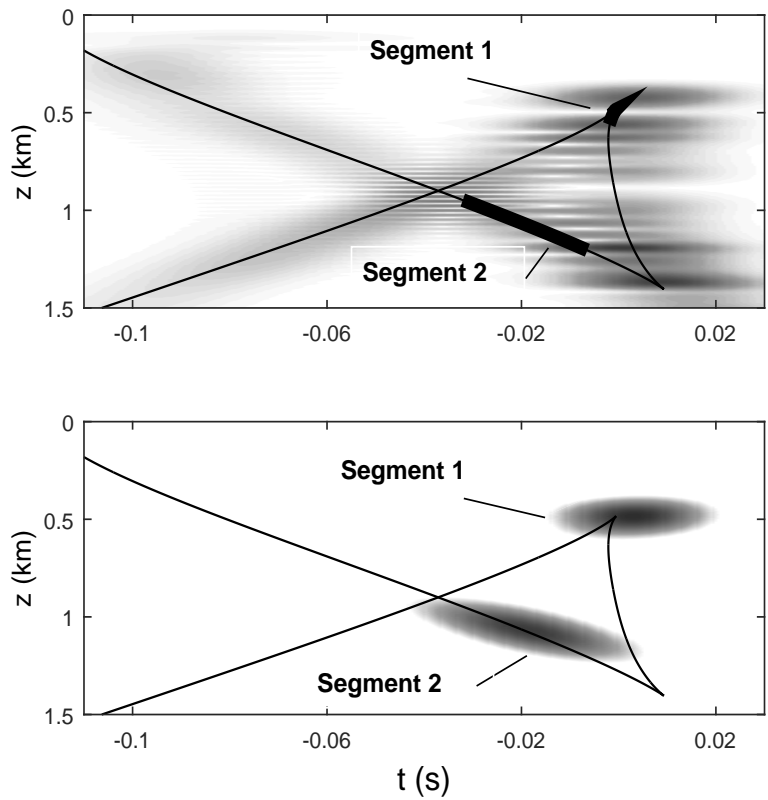}
\end{center}
\caption{Amplitudes of the total field (upper panel) and its
stable components associated with Segments 1 and 2 (lower panel)
in the plane time -- depth. The same thin line in both panels
depicts the timefront. Parts of the timefront in the upper panel
depicting
arrivals of rays forming Segments 1 and 2 are shown in bold.}%
\label{fig10}%
\end{figure}

Upper panel in Fig. \ref{fig10} presents the distribution of field
amplitude $\left\vert v_{n}\left(  z,t\right)  \right\vert $ in
the plane time -- depth for a particular realization of sound
speed perturbation. Thin solid lines in both panels depict the
timefront representing the ray arrival depth as a function of the
ray travel time. Ray travel times are reckoned from $r/c$. Pieces
of the timefront formed by rays belonging to Segments 1 and 2 are
shown in bold. Lower panel represents amplitudes of CS components
$\left\vert v_{a}\left( z,t\right)  \right\vert $ associated with
the Segments. They are localized near the corresponding parts of
the timefronts. Distributions of amplitudes $\left\vert
v_{b}\left( z,t\right) \right\vert $ representing modal components
look similarly (not shown).

Stable components $v_{a}\left(  z,t\right)  $ associated with Segments 1 and 2
are weakly sensitive to perturbation $\delta c$: similarity coefficient
$B_{\text{tr}}\left(  v_{a,n},v_{a,1}\right)  $ determined by Eq. (\ref{Bt})
for both segments exceed 0.9 for all $n=1,\ldots,22$. The same is true for
coefficients $B_{\text{tr}}\left(  v_{b,n},v_{b,1}\right)  $.

\begin{figure}[ptb]
\begin{center}
\includegraphics[
height=2.6411in, width=4.5189in ]{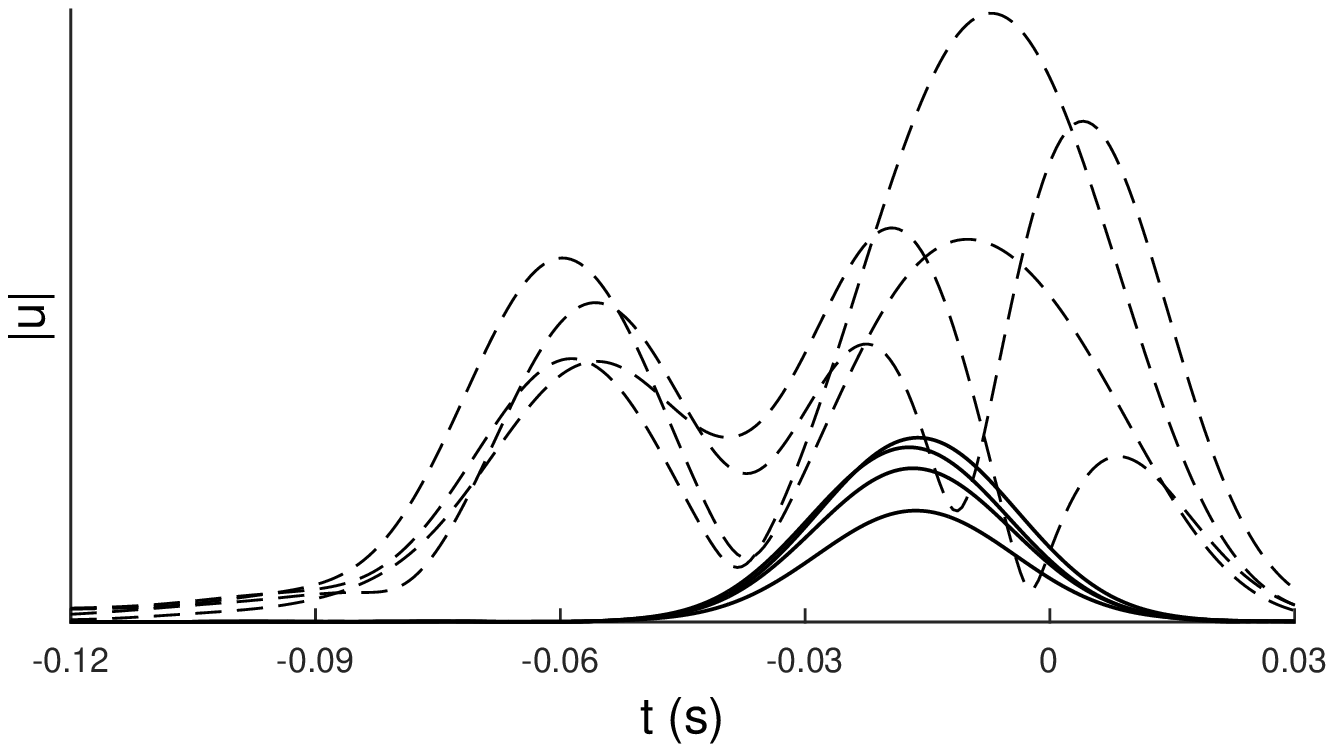}
\end{center}
\caption{Sound pulses (dashed lines) at depth $z_{c}=1.08$ km and
their stable components associated with Segment 2 (solid
lines) in four realizations of random waveguide.}%
\label{fig11}%
\end{figure}

Sound pulse $v_{a}\left(  z,t\right)  $ at a fixed depth $z$ can
be interpreted as a stable component of signal $v\left( z,t\right)
$, recorded by a single hydrophone. Solid lines in Fig.
\ref{fig11} show pulses $\left\vert v_{a,n}\left( z_{c},t\right)
\right\vert $ at $z_{c}=1.08$ km (central depth of Segment 2) for
four realizations of perturbation $\delta c$. It is seen that
these four pulses are indeed more similar to each other than total
signals $\left\vert v_{n}\left( z_{c},t\right) \right\vert $
(dashed lines) arriving at the same depth in the same realizations
of the waveguide.

In order to quantify the closeness of pulses $v(z_{с },t)$ and $\tilde{v}(z_{с
},t)$ arriving at the same depths $z_{с }$ in different realizations of the
random waveguide, by analogy with Eq. (\ref{Bt}), we will use the similarity
coefficient%
\[
B\left(  v,\tilde{v}\right)  =\max_{\tau}\frac{\left\vert \int dt~v\left(
z_{c},t\right)  \tilde{v}_{c}^{\ast}\left(  z_{c},t+\tau\right)  \right\vert
}{\left(  \int dt~\left\vert v(z_{c},t\right\vert ^{2}\right)  ^{1/2}\left(
\int dt~\left\vert \tilde{v}(z_{c},t\right\vert ^{2}\right)  ^{1/2}}%
\]%
\begin{equation}
=\max_{\tau}\frac{\left\vert \int df~u\left(  z_{c},f\right)  \tilde{u}^{\ast
}\left(  z_{c},f\right)  \left\vert g\left(  f\right)  \right\vert ^{2}e^{2\pi
if\tau}\right\vert }{\left(  \int df~\left\vert u(z_{c},f\right\vert
^{2}\left\vert g\left(  f\right)  \right\vert ^{2}\right)  ^{1/2}\left(  \int
df~\left\vert \tilde{u}(z_{c},f\right\vert ^{2}\left\vert g\left(  f\right)
\right\vert ^{2}\right)  ^{1/2}}. \label{Bb}%
\end{equation}

\begin{figure}[ptb]
\begin{center}
\includegraphics[
height=3.5in, width=4.6in ]{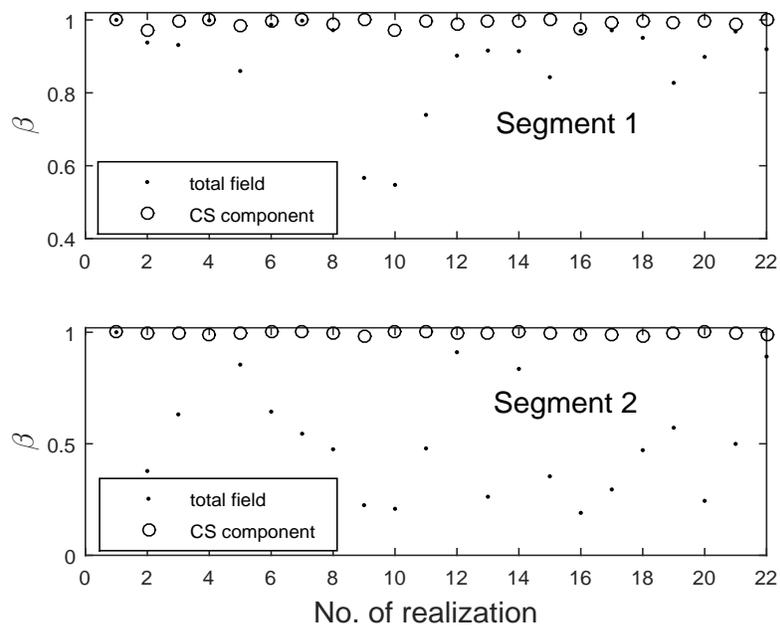}
\end{center}
\caption{Similarity coefficients of signals (points) and
their stable components (circles) in the perturbed and unperturbed waveguide
at central depths of Segment 1 (upper panel) and Segment 2 (lower panel).}%
\label{fig12}%
\end{figure}

Upper panel in Fig. \ref{fig12} depicts similarity coefficients
$B\left( v_{a,n},v_{a,1}\right)  $ of CS components of pulses
associated with Segment 1 and received at the central depth of
this segment $z_{c}=0.49$ km (circles). Points show similarity
coefficients of total signals received at the same horizon. Lower
panel presents similar results obtained for CS components
associated with Segment 2 and for total signals received at the
central depth of this segment $z_{c}=1.08$ km. Numerical
simulation confirms stability of pulses associated with both
segments. A similar result was obtained for modal components
expressed by functions $v_{b,n}$ (not shown).

\section{Conclusion \label{sec:concl}}

The procedure proposed in this paper for evaluating components of
the wave field which are stable with respect to small variations
of the sound speed field includes the following steps. (i) A dense
fan of rays escaping the source with different starting momenta
$p_{0}$ is traced in the unperturbed waveguide. (ii) On the basis
of available a priori information, an ensemble of waveguides with
admissible realizations of the sound speed field is constructed.
This ensemble can model not only random sound speed flucations but
the uncertainty in parameters of the regular waveguide as well.
(iii) Using Eq. (\ref{dS}), eikonal increments $\delta S$ at the
observation range are evaluated for every fan ray in every
realization of the waveguide. Then, for any given frequency $f$
and for each starting momentum $p_{0}$ such an interval $\delta
p_{0}$ is evaluated that phase increments $k\delta S$ of rays with
starting momenta from interval $p_{0}\pm\delta p_{0}/2$ are close
and these rays form a stable component of the total field. (iv)
This stable component is constructed from coherent states by
formulas (\ref{us-coh}) -- (\ref{va-ua}), or from Brillouin waves
by formulas (\ref{us-mode}) and (\ref{Xi-mode}). Application of
this procedure is illustrated with the numerical example. We
assume that a similar procedure can be applied for finding field
components stable with respect to variations of the waveguide
boundary.

The weak point of our approach is the fact that at present it has
no rigorous justification. Stability of field components expressed
by Eqs. (\ref{us-coh}) and (\ref{us-mode}), is actually only our
conjecture. It is based on simple heuristic arguments formulated
in Sec. \ref{sec:heuristic} and supported by results of numerical
simulation. Note that the stable component associated with Segment
1 is defined in the waveguide area with caustic (see Fig.
\ref{fig4}), where the geometrical optics approximation fails.
This and other examples not presented in this paper suggest that
the approach under consideration is applicable at caustics.

Numerical simulation shows that, in spite of our assumption made
in Sec. \ref{sec:heuristic}, the amplitudes of stable components
in different realizations of random waveguide may significantly
differ (see Figs. \ref{fig6} and \ref{fig11}). Weak sensitivity of
functions $u_{a,b}$ and $v_{a,b}$ to sound speed fluctuations is
manifested in the fact that the similarity coefficients of these
components in the perturbed and unperturbed waveguide are close to
unity.

The lack of rigorous justification complicates the optimal choice of scales
$\Delta_{z}$ and $\Delta_{p}$ in the calculation of the CS component and scale
$\Delta_{g}$ in the calculation of the modal component. Moreover, the proposed
methods of constructing stable components may not be optimal. For example,
when calculating the modal component, it may make sense to consider
$\Delta_{g}$, the effective number of added Brillouin waves, as a function of
depth. The relationship between two representations of stable components given
by Eqs. (\ref{us-coh}) and (\ref{us-mode}) is not yet studied. These issues
require a further investigation which we plan to perform in next publications.

Another group of questions which is not addressed in the present paper is
connected with the practical measurement of stable components (requirements
for the receiving antenna) and their possible use in solving different
problems of underwater acoustics (source localization, remote sensing, noise
interferometry, etc). These issues, we also plan to consider elsewhere.

\section*{Acknowledgment}

The work was supported by the Program ``Fundamentals of acoustic
diagnostics of artificial and natural media'' of Physical Sciences
Division of Russian Academy of Sciences, and Grants No.
15-02-04042 and 15-42-02390 from the Russian Foundation for Basic
Research.


\begin{thebibliography}{99}                                                                                               %
\bibitem{Dowling2008}
K.R. James and D.R. Dowling, ``A method for approximating
acoustic-field-amplitude uncertainty caused by environmental
uncertainties'', J. Acoust. Soc. Am., \textbf{124}, 1465-–1476
(2008).

\bibitem{LePage2006}
K.D. LePage, ``Modeling propagation and reverberation sensitivity
to oceanographic and seabed variability'', IEEE J. Ocean Eng.,
\textbf{31}, 402-–412 (2006).

\bibitem{Gerstoft1998}
P. Gerstoft and C.F. Mecklenbr\"{a}uker, ``Ocean acoustic
inversion with estimation of a posteriori probability
distributions'', J. Acoust. Soc. Am., \textbf{104}, 808-–819
(1998).

\bibitem{Lin2006}
Y.-T.Lin, C.-F. Chen, and J.F. Lynch, ``An equivalent transform
method for evaluating the effect of water-column mismatch on
geoacoustic inversion'', J. Ocean. Eng., \textbf{31}, 284--298
(2006).

\bibitem {FDMWZ79}S.M. Flatte, R.~Dashen, W.M. Munk, K.M. Watson, and
F.~Zakhariasen, \emph{Sound transmission through a fluctuating
ocean} (Cambringe U.P., London, 1979), Chaps. 7, 8, 11.

\bibitem {BL2003}L.M. Brekhovskikh and Yu.P. Lysanov, \emph{Fundamentals of
Ocean Acoustics} (Springer-Verlag, New York, 2003), Chaps. 6, 10.

\bibitem {Vbook2010}D.~Makarov, S.~Prants, A.~Virovlyansky, and G.~Zaslavsky.
\emph{Ray and wave chaos in ocean acoustics} (Word Scientific, New
Jersey, 2010), pp. 22-25.

\bibitem{V2003c}
F.J. Beron-Vera, M.G. Brown, J.A. Colosi, S.~Tomsovic, A.L.
Virovlyansky, M.A. Wolfson, and G.M. Zaslavsky, ``Ray dynamics in
a long-range acoustic propagation experiment'', J. Acoust. Soc.
Am., \textbf{114}, 1226--1242 (2003).

\bibitem {CB98}
J.A. Colosi and M.G. Brown, ``Efficient numerical simulation of
stochastic internal-wave-induced sound-speed perturbation field'',
J. Acoust. Soc. Am., \textbf{103}, 2232--2235 (1998).

\bibitem {CM68}
P. Carruthers and M.M. Nieto, ``Phase and angle variables in
quantum mechanics'', Rev. Mod. Phys., \textbf(40), 411--440
(1968).

\bibitem {JKPS2011}F.B. Jensen, W.A. Kuperman, M.B. Porter, and H.~Schmidt,
\emph{Computational Ocean Acoustics} (Springer, New York, 2011),
Chaps. 5,6.

\bibitem {BG99}L.M. Brekhovskikh and O.A. Godin.
\emph{ Acoustics of Layered Media. II: Point Sources and Bounded
Beams} (Springer-Verlag, Berlin, 1999), Chap. 7.

\bibitem {LLquant}L.D. Landau and E.M. Lifshitz.
\emph{ Quantum mechanics} (Pergamon Press, Oxford, 1977), Chaps.
2, 7.

\end{thebibliography}
\end{document}